\documentclass[11pt,onecolumn,a4paper,journal]{IEEEtran}
\IEEEoverridecommandlockouts

\usepackage{cite}
\usepackage{amsmath,amssymb,amsfonts}

\newcommand{\vect}[1]{\boldsymbol{#1}}
\newcommand{\mat}[1]{\boldsymbol{#1}}
\newcommand{\bpsi}{\boldsymbol{\psi}}
\providecommand{\E}{\ensuremath{\mathbb{E}}}
\providecommand{\C}{\ensuremath{\mathbb{C}}}
\usepackage{graphicx}
\usepackage{booktabs}
\usepackage[nomain, acronym, nonumberlist]{glossaries}
\newacronym{ape}{APE}{active pilot eavesdropper}
\newacronym{aes}{AES}{advanced encryption standard}
\newacronym{an}{AN}{artificial noise}
\newacronym{ap}{AP}{access point}
\newacronym{bs}{BS}{base station}
\newacronym{cdf}{CDF}{cumulative distribution function}
\newacronym{cf}{CF}{cell-free}
\newacronym{cnn}{CNN}{convolutional neural network}
\newacronym{cpu}{CPU}{central processing unit}
\newacronym{flop}{FLOP}{floating-point operation}
\newacronym{gnn}{GNN}{graph neural network}
\newacronym{ibo}{IBO}{input back-off}
\newacronym{los}{LoS}{line-of-sight}
\newacronym{mimo}{MIMO}{multiple-input multiple-output}
\newacronym{miso}{MISO}{multiple-input single-output}
\newacronym{mlp}{MLP}{multilayer perceptron}
\newacronym{mmse}{MMSE}{minimum mean square error}
\newacronym{nlos}{NLoS}{non-line-of-sight}
\newacronym{pa}{PA}{power amplifier}
\newacronym{pls}{PLS}{physical layer security}
\newacronym{sindr}{SINDR}{signal-to-interference-noise-and-distortion ratio}
\newacronym{svd}{SVD}{singular value decomposition}
\newacronym{tdd}{TDD}{time division duplexing}
\newacronym{tls}{TLS}{transport layer security}
\newacronym{z3ro}{Z3RO}{zero third-order distortion}
\newacronym{zf}{ZF}{zero-forcing}

\makeatletter
\long\def\@makecaption#1#2{
  \small
  \vskip\abovecaptionskip
  \sbox\@tempboxa{#1.\nobreakspace#2}
  \ifdim \wd\@tempboxa >\hsize
    #1.\nobreakspace#2\par
  \else
    \hb@xt@\hsize{\hfil\box\@tempboxa\hfil}
  \fi
  \vskip\belowcaptionskip}
\makeatother

\graphicspath{{figures/}}

\begin{document}

\title{Secure Distortion-Aware GNN Precoding under
       Active Eavesdropping in Cell-Free Networks}

\author{Reza Ghasemi Alavicheh,
Thomas Feys,
Md Arifur Rahman,
Fran\c{c}ois Rottenberg
\thanks{R. Ghasemi Alavicheh is with the Research and Innovation Department, IS-Wireless, Piaseczno, Poland (e-mail: r.ghasemi@is-wireless.com), and also with ESAT-DRAMCO, KU Leuven, Ghent, Belgium (e-mail: reza.ghasemialavicheh@kuleuven.be).}
\thanks{T. Feys and F. Rottenberg are with ESAT-DRAMCO, KU Leuven, Ghent, Belgium (e-mail: \{thomas.feys, francois.rottenberg\}@kuleuven.be).}
\thanks{M. A. Rahman is with the Research and Innovation Department, IS-Wireless, Piaseczno, Poland (e-mail: a.rahman@is-wireless.com).}
\thanks{This work has received funding from the European Union's Horizon Europe research and innovation programme under the Marie Sk{\l}odowska-Curie grant agreement No. 101120332 (EMPOWER-6G project).}}

\maketitle

\begin{abstract}
This paper considers a \gls{cf} network under active eavesdropping. We show that nonlinear \gls{pa} distortion, together with the per-\gls{ap} large-scale-gain fingerprint of the attack, can be exploited across cooperating \glspl{ap} to enhance \gls{pls}. Classical \gls{zf}-\gls{an} precoders rely on a linear transmitter and an unbiased estimate, and thus cannot exploit either. We propose a distortion-aware \gls{gnn} precoder that designs the signal beams at cooperating \glspl{ap} and exploits both the \gls{pa} distortion and the per-\gls{ap} fingerprint to enhance the secrecy rate, combined with a null-space \gls{an} block to form the \gls{gnn}-\gls{an} scheme. The \gls{gnn} is trained on channel estimates biased by an uplink pilot-spoofing attack to maximise the average per-user secrecy rate. Simulations across different numbers of \glspl{ap} show uniformly higher outage secrecy rates than \gls{zf}-\gls{an} under attack, reaching gains of $15\%$ at the $1\%$ outage level and $8.9\%$ at the $5\%$ outage level at $L=32$ \glspl{ap}, \gls{ibo} $= -1$ dB, and attacker-to-targeted user pilot power ratio of $0.5$, under the same signal/\gls{an} power split.
\end{abstract}

\begin{IEEEkeywords}
Physical layer security, cell-free networks, graph neural network, artificial noise, pilot spoofing, active eavesdropper, power amplifier nonlinearity, secrecy outage.
\end{IEEEkeywords}

\glsresetall

\section{Introduction}
Modern wireless networks target a wide range of services, from high-throughput mobile broadband and mission-critical low-latency links to massive device connectivity~\cite{noauthor_guidelines_nodate}. All of these carry private and sensitive data, so the security of the wireless link matters in every deployment scenario. As more of these services move over the air, the cost of a single leaked message grows in the same proportion. Modern systems address this challenge with encryption at the upper layers of the protocol stack. Standardised solutions such as the \gls{aes} and \gls{tls} are mature, widely deployed, and independent of the wireless channel. However, they rest on computational hardness assumptions, are sensitive to advances in computing power, and require careful key management~\cite{poor_wireless_2017}.

\Gls{pls} is instead built into the signal itself, rooted in information theory~\cite{wyner_wire-tap_1975}, and independent of the eavesdropper's computing capability. It designs the transmit signal so that legitimate users see a strong effective channel while an eavesdropper sees stronger noise. Realising \gls{pls} in a \gls{cf} network is, however, complicated by hardware impairments such as phase noise, in-phase/quadrature (I/Q) imbalance, and \gls{pa} nonlinearity; the \gls{pa} dominates because it is driven close to saturation for energy efficiency, producing distortion in the transmit signal~\cite{rapp_effects_1991,moghadam_energy_2018}. Distortion-aware precoders cancel this distortion around legitimate users~\cite{aghdam_distortion-aware_2021,rottenberg_z3ro_2023}, yet a learned precoder can use the same distortion as a secrecy resource against an eavesdropper. Beyond passive eavesdropping, \gls{pls} also faces active attacks. In \gls{cf} networks, downlink precoding relies on uplink pilots for channel estimation~\cite{ngo_cell-free_2017}; an active eavesdropper can inject a spoofed pilot to bias the estimate, tilting the beam toward them and leaving a per-\gls{ap} signature on it. Unlike prior works on distortion-aware secure precoding~\cite{alavicheh_leveraging_2025,alavicheh_when_2026}, which target single-cell wiretap channels with a passive eavesdropper, the joint handling of \gls{ap} cooperation, spoofed channel estimates, and nonlinear \gls{pa} distortion in this setting is unexplored and challenging, motivating more advanced precoder designs.

\subsection{State of the Art}
\label{sec:sota}
The prior work most relevant to this paper can be organised into three lines, namely secure precoding under active eavesdropping, distortion-aware precoding under nonlinear \glspl{pa}, and learning-based precoding with \glspl{gnn}. 

On the first, pilot spoofing during uplink training, which is an active security scenario considered in this work, has been analysed both in cellular massive \gls{mimo}~\cite{zhou_pilot_2012,kapetanovic_physical_2015,wu_secure_2016} and, more recently, in \gls{cf} deployments~\cite{atiya_secure_2024,hoang_cell-free_2018}. The standard tool for enforcing secrecy at the transmit side is \gls{an}, which spends part of the transmit power on directions orthogonal to the legitimate channels~\cite{goel_guaranteeing_2008}; combining \gls{an} with a \gls{zf} beamformer yields the widely used ZF-\gls{an} precoder~\cite{zhu_linear_2016}. Under the pilot spoofing attack, AN-aided precoders have been proposed to jointly optimise the power split between the information signal and the \gls{an} based on the contaminated channel estimate~\cite{wu_secure_2016}, while a \gls{cf} extension formulated secrecy-rate maximisation as a downlink power allocation problem across \glspl{ap}~\cite{hoang_cell-free_2018}, and a subsequent work characterised how residual hardware impairments impact the achievable secrecy rate in the same setting~\cite{zhang_secure_2020}; more recent work has continued to develop power-control and \gls{ap}-cooperation strategies in the presence of active eavesdroppers~\cite{atiya_secure_2024}. However, none of these works consider a realistic \gls{pa} model and exploit hardware distortion as a secrecy resource. 

On the second, exploiting \gls{pa} distortion for secrecy was recently proposed by showing that the same distortion can be redirected toward the eavesdropper~\cite{alavicheh_leveraging_2025}. This idea was subsequently generalised in~\cite{alavicheh_when_2026}. Both of these studies, however, were carried out in a cellular multi-user \gls{miso} setting with a passive eavesdropper. 

On the third, \glspl{gnn} are well suited to precoding because they respect the permutation equivariance between users and between \glspl{ap}~\cite{scarselli_graph_2009,shen_graph_2021}. In this context, \gls{gnn} precoders have already outperformed classical baselines in the single-cell \gls{pa}-nonlinear setting~\cite{feys_toward_2025} and have been applied to the \gls{cf} downlink to coordinate multiple \glspl{ap}~\cite{salaun_gnn_2022}, while broader learning-based methods have addressed precoding-aware resource allocation in \gls{cf} networks~\cite{girycki_learning-based_2024}. Taken together, the three lines above cover active eavesdropping like pilot spoofing, distortion-aware precoding, and \gls{gnn}-based precoding separately, but none of them considers a \gls{cf} downlink under active pilot spoofing with a nonlinear \gls{pa} model in which a learned precoder exploits both the distortion and the per-\gls{ap} fingerprint of the attack, shaped with an \gls{an} block sharing the same power budget as the \gls{zf}-\gls{an} baseline.

\subsection{Contributions}
\label{sec:contributions}
To close this gap, we propose a \gls{gnn}-\gls{an} precoder that jointly designs the signal beams at cooperating \glspl{ap} under a nonlinear \gls{pa} model to enhance the secrecy rate under active pilot spoofing.
Against this background, our contributions are as follows.
\begin{itemize}
    \item We formulate the secure downlink of a \gls{cf} network with nonlinear polynomial \glspl{pa} under an active pilot-spoofing eavesdropper.
   \item We propose a distortion-aware \gls{gnn} precoder that jointly generates the signal beams across cooperating \glspl{ap}, complemented by a null-space \gls{an} block, and trained to maximise the average per-user secrecy rate; we benchmark it against ZF-\gls{an}~\cite{zhu_linear_2016}.
    \item Under the same \gls{an} block as ZF-\gls{an}, the scheme improves the $1\%$ and $5\%$ outage secrecy rates.
\end{itemize}

Bold lower/uppercase letters denote vectors/matrices; $\mat{I}$ the identity, $(\cdot)^H,(\cdot)^*$ Hermitian and complex conjugate, $\odot$ Hadamard product, $\E[\cdot]$ expectation, $\mathcal{CN}(\mu,\sigma^2)$ complex Gaussian, $[x]^+=\max(0,x)$.
The rest of the paper is organised as follows. Section~\ref{sec:system_model} presents the system model and problem formulation. Section~\ref{sec:gnn_an} describes the proposed \gls{gnn}-\gls{an} precoder and its training. Section~\ref{sec:results} reports the numerical results, and Section~\ref{sec:conclusion} concludes the paper.

\section{System Model and Problem Formulation}
\label{sec:system_model}

\begin{figure}[t]
    \centering
    \includegraphics[width=0.65\linewidth]{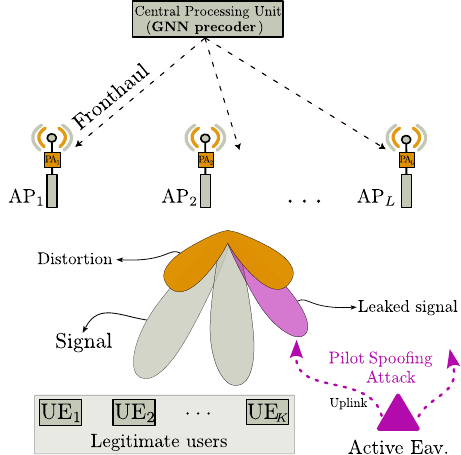}
    \caption{Cell-free downlink with active pilot spoofing in the uplink.}
    \label{fig:system_model}
\end{figure}

Fig.~\ref{fig:system_model} shows the considered \gls{cf} downlink, with $L$ single-antenna \glspl{ap}, $K$ single-antenna legitimate users, and a single \gls{ape} equipped with one antenna. All \glspl{ap} are connected to a \gls{cpu} through an ideal fronthaul and cooperate through the proposed precoder. The channel from \gls{ap} $l$ to user $k$ is modelled as independent Rayleigh block fading
\begin{equation}
    h_{l,k} = \sqrt{\beta_{l,k}}\, f_{l,k}, \qquad f_{l,k} \sim \mathcal{CN}(0,1)
    \label{eq:channel}
\end{equation}
where $\beta_{l,k}$ is the large-scale gain, following the ITU-R 3D UMi Street-Canyon path-loss model with a \gls{los} and \gls{nlos} profile and log-normal shadowing~\cite{noauthor_guidelines_nodate}, and $f_{l,k}$ is the small-scale coefficient, constant within a coherence block and independent across blocks. The channel of the eavesdropper, hereafter referred to as Eve, $h_{l,e}$ obeys the same law with parameter $\beta_{l,e}$.

\subsection{System Model}
The \glspl{ap} obtain the downlink channel through a $\tau_p$-symbol uplink pilot phase under \gls{tdd} channel reciprocity. Denoting by $\bpsi_k\!\in\!\C^{\tau_p}$ the pilot of user $k$, our formulation supports orthogonal pilot families satisfying $\bpsi_k^H\bpsi_{k'}=\tau_p\delta_{k,k'}$, where $\delta_{k,k'}$ is the Kronecker delta; we assume cyclic-shifted Zadoff--Chu sequences of prime length $\tau_p\!\geq\!K$ with a single root, so the different users' pilots are mutually orthogonal~\cite{hyder_zadoffchu_2017}. The pilot codebook is standardised and hence known to any eavesdropper. During the pilot training phase, the \gls{ape} transmits the pilot of a target user $p$ in the classical pilot spoofing setup of~\cite{zhou_pilot_2012,kapetanovic_physical_2015,wu_secure_2016}, so the pilot observation at \gls{ap} $l$ is
\begin{equation}
    \vect{y}_l^{\mathrm{p}} = \sum_{k=1}^{K}\sqrt{q_k}\,h_{l,k}\,\bpsi_k
      + \sqrt{q_e}\,h_{l,e}\,\bpsi_p + \vect{n}_l
    \label{eq:pilot_rx}
\end{equation}
where $\vect{y}_l^{\mathrm{p}},\vect{n}_l\!\in\!\C^{\tau_p}$, $q_k$ is the pilot power of user $k$ and $\beta_{l,k}$ the corresponding large-scale gain (so $q_p$ and $\beta_{l,p}$ denote the pilot power and large-scale gain of the target user $p$), $q_e$ is the attacker pilot power, and $\vect{n}_l\!\sim\!\mathcal{CN}(\vect{0},\sigma^2\mat{I}_{\tau_p})$ with $\sigma^2$ the uplink noise variance and $\mat{I}_{\tau_p}$ the $\tau_p\!\times\!\tau_p$ identity matrix, so the noise samples are uncorrelated across pilot symbols. 

De-spreading with $y_{l,p}\!=\!\bpsi_p^H\vect{y}_l^{\mathrm{p}}/\tau_p$ and applying the \gls{mmse} estimator $\hat h_{l,p}\!=\!\bigl(\E[h_{l,p}\,y_{l,p}^*]/\E[|y_{l,p}|^2]\bigr)\,y_{l,p}$ following~\cite{wu_secure_2016,hoang_cell-free_2018,atiya_secure_2024} yields
\begin{equation}
    \hat h_{l,p}^{\mathrm{att}} = \frac{\sqrt{q_p}\beta_{l,p}}{\Psi_{l,p}}\,y_{l,p},
    \;
    \Psi_{l,p} = q_p\beta_{l,p} + q_e\beta_{l,e} + \tfrac{\sigma^2}{\tau_p}.
    \label{eq:mmse_att}
\end{equation}
The extra term $q_e\beta_{l,e}$ in $\Psi_{l,p}$ biases the estimate toward the attacker channel; a precoder derived analytically from this spoofed estimate $\hat h_{l,p}^{\mathrm{att}}$, where the superscript $\mathrm{att}$ marks the attack scenario and is applied to all user estimates for uniformity, therefore leaks the intended beam toward the \gls{ape}~\cite{kapetanovic_physical_2015}. Estimates $\hat h_{l,k}^{\mathrm{att}}$ for $k\!\neq\!p$ follow the same form without the $q_e\beta_{l,e}$ term, since the attacker mimics the pilot of one target user $p$ at any given time following the classical spoofing model of~\cite{zhou_pilot_2012,kapetanovic_physical_2015}. The large-scale statistics of the legitimate users are treated as known at the \gls{cpu}~\cite{ngo_cell-free_2017,hoang_cell-free_2018,wu_secure_2016}, obtainable in practice from long-term averaging over many coherence blocks.

Given the biased estimates, the \gls{cpu} assembles a per-\gls{ap}, per-user signal precoder scalar $w_{l,k}\!\in\!\C$, gathered across \glspl{ap} into the vector $\vect{w}_k=[w_{1,k},\dots,w_{L,k}]^T\!\in\!\C^L$, and stacked across users into $\mat{W}_{\mathrm{sig}}=[\vect{w}_1,\dots,\vect{w}_K]\!\in\!\C^{L\times K}$, together with a null-space \gls{an} component $z_l\!\in\!\C$; \gls{ap} $l$ then transmits $x_l = \sum_{k} w_{l,k}\,s_k + z_l$ with i.i.d.\ data symbols $s_k\!\sim\!\mathcal{CN}(0,1)$. Similarly, collecting the \gls{an} components into $\vect{z}=[z_1,\dots,z_L]^T$, we write $\vect{z} = \mat{W}_{\mathrm{an}}{\boldsymbol{\eta}}$ with ${\boldsymbol{\eta}}\!\in\!\C^{L-K}$, ${\boldsymbol{\eta}}\!\sim\!\mathcal{CN}(\vect{0},\mat{I})$, and the columns of $\mat{W}_{\mathrm{an}}\!\in\!\C^{L\times(L-K)}$ spanning $\mathrm{null}(\hat{\mat{H}}^H)$, where $\hat{\mat{H}}\!\in\!\C^{L\times K}$ collects the spoofed estimates $\hat h_{l,k}^{\mathrm{att}}$; the scaling of $\mat{W}_{\mathrm{an}}$ is specified in Section~\ref{sec:gnn_an}, \eqref{eq:an}. The per-\gls{ap} power constraint $\E[|x_l|^2]\!\leq\!P_l^{\max}$ is enforced on the full transmit signal, so the mean transmit power and the distortion covariance introduced by the Bussgang decomposition below both include the signal and \gls{an} components.

Each \gls{ap} then applies a memoryless nonlinear \gls{pa} $\tilde x_l=f_{\mathrm{PA}}(x_l)$. We adopt an odd-order polynomial obtained by ridge regression of a Rapp model~\cite{rapp_effects_1991} at the target \gls{ibo}, defined as $\mathrm{IBO}\,\text{(dB)}=10\log_{10}(P_{\mathrm{in}}/P_{\mathrm{sat}})$ with $P_{\mathrm{in}}=\E[|x_l|^2]$,
\begin{equation}
    \tilde x_l = \sum_{n=0}^{N}b_{2n+1}\,x_l\,|x_l|^{2n}
    \label{eq:pa_poly}
\end{equation}
fitted using the odd basis terms up to degree $2N{+}1$, with ridge regularisation. Bussgang's theorem~\cite{aghdam_distortion-aware_2021,moghadam_energy_2018,alavicheh_leveraging_2025} then decomposes~\eqref{eq:pa_poly} as $\tilde x_l=g_l\,x_l+d_l$ with $\E[d_l x_l^*]=0$, where the linear coefficient $g_l=\sum_{n=0}^{N}(n{+}1)!\,b_{2n+1}\,\bigl(\E[|x_l|^2]\bigr)^n$ depends only on the transmit power $\E[|x_l|^2]$, $d_l$ is the residual per-\gls{ap} distortion noise uncorrelated with $x_l$, and the distortion covariance $\mat{C}_d\!=\!\E[\vect{d}\vect{d}^H]\!=\!\sum_{n=1}^{N}\mat{L}_n(\mat{C}_x\!\odot\!|\mat{C}_x|^{2n})\mat{L}_n^H$, with $\mat{C}_x\!=\!\E[\vect{x}\vect{x}^H]$ the transmit covariance and $\mat{L}_n\!=\!\tfrac{1}{\sqrt{n+1}}\sum_{m=n}^{N}\binom{m}{n}(m{+}1)!\,b_{2m+1}\,\mathrm{diag}(\mat{C}_x)^{m-n}$~\cite{aghdam_distortion-aware_2021}, is non-diagonal because the transmit signals of different \glspl{ap} share the data symbols $\{s_k\}$~\cite{alavicheh_when_2026}.

Collecting $\vect{g}=[g_1,\dots,g_L]^T$ and $\vect{d}=[d_1,\dots,d_L]^T$, the signal at user $k$ is $y_k=\sum_l h_{l,k}^*(g_l x_l+d_l)+n_k$ with $n_k\!\sim\!\mathcal{CN}(0,\sigma_n^2)$; the eavesdropper is assumed to have the same receive-noise variance $\sigma_n^2$. The \gls{sindr} at user $k$ and at Eve on stream $k$ are, respectively,
\begin{align}
    \gamma_k     &= \frac{\bigl|\vect{h}_k^H\mat{G}\vect{w}_k\bigr|^2}{\sum_{j\neq k}\bigl|\vect{h}_k^H\mat{G}\vect{w}_j\bigr|^2 + \vect{h}_k^H\mat{C}_{\mathrm{an}}\vect{h}_k + \vect{h}_k^H\mat{C}_d\vect{h}_k + \sigma_n^2} \label{eq:sindr_leg}\\
    \gamma_{e,k} &= \frac{\bigl|\vect{h}_e^H\mat{G}\vect{w}_k\bigr|^2}{\vect{h}_e^H\mat{C}_{\mathrm{an}}\vect{h}_e + \vect{h}_e^H\mat{C}_d\vect{h}_e + \sigma_n^2} \label{eq:sindr_eve}
\end{align}
where $\vect{h}_k\!\in\!\C^L$ contains the channels from all \glspl{ap} to user $k$, similarly for $\vect{h}_e$, $\mat{G}\!=\!\mathrm{diag}(\vect{g})\!\in\!\C^{L\times L}$, and $\mat{C}_{\mathrm{an}}\!=\!\mat{G}\mat{W}_{\mathrm{an}}\mat{W}_{\mathrm{an}}^H\mat{G}^H\!\in\!\C^{L\times L}$ is the \gls{an} covariance at the receiver; $\vect{h}_k^H\mat{G}\vect{w}_k\!=\!\sum_l h_{l,k}^* g_l w_{l,k}$ captures the coherent combining across \glspl{ap} and $\sum_{j\neq k}\bigl|\vect{h}_k^H\mat{G}\vect{w}_j\bigr|^2$ is the inter-user interference at user $k$. The \gls{an} leakage at legitimate users is nonzero for two reasons, first, because $\mat{W}_{\mathrm{an}}$ is designed on the spoofed estimate $\hat{\mat{H}}$ rather than on the true channel, and second, because the per-\gls{ap} Bussgang gains $g_l$ break the null-space alignment through $\mat{G}$. Equation~\eqref{eq:sindr_eve} gives Eve's rate on stream $k$ under a challenging scenario for the transmitter, in which Eve suffers no inter-user interference and is limited only by \gls{an}, distortion, and noise~\cite{zhou_secure_2010}; pilot spoofing biases only the target stream $k=p$, but~\eqref{eq:sindr_eve} is evaluated for every $k$ to give each user its own secrecy rate.

\subsection{Problem Formulation}
The instantaneous per-user secrecy rate is $R_{\mathrm{sec},k}=[\log_2(1+\gamma_k)-\log_2(1+\gamma_{e,k})]^+$~\cite{wyner_wire-tap_1975}, and the design problem is
\begin{equation}
    \max_{\{w_{l,k}\}}\ \E_{\vect{h}_e}\!\Bigl[\tfrac{1}{K}\textstyle\sum_{k=1}^{K}R_{\mathrm{sec},k}\Bigr]\ \ \text{s.t.}\ \ \E[|x_l|^2]\!\leq\!P_l^{\max}.
    \label{eq:problem}
\end{equation}
Problem~\eqref{eq:problem} is non-convex and its variable dimension grows with $LK$; the nested \gls{sindr} terms in the log-difference secrecy rate, together with the nonlinear \gls{pa} distortion shared across all \glspl{ap}, are difficult to address directly with standard convexification, motivating the learning-based precoder below.
\section{GNN-AN Precoder for Secrecy}
\label{sec:gnn_an}

\begin{figure}[!t]
    \centering
    \includegraphics[width=0.7\columnwidth]{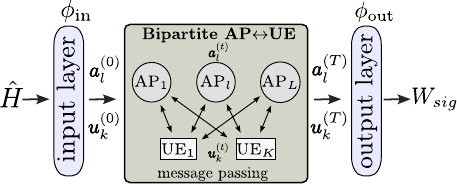}
    \caption{\gls{gnn} precoder $f_\theta$ with $T$ rounds of message passing.}
    \label{fig:gnn_arch}
\end{figure}

Problem~\eqref{eq:problem} is invariant to any joint permutation of the user indices, so reordering users in the channel matrix must produce the same reordering of the precoder columns. Standard \glspl{mlp} and \glspl{cnn} applied to the channel do not naturally respect this symmetry and would have to learn it from data, whereas a \gls{gnn} whose graph mirrors the antenna-user structure enforces it by design~\cite{scarselli_graph_2009,shen_graph_2021}, using about $100\times$ fewer trainable parameters than \gls{cnn} and \gls{mlp} baselines on a related precoding task~\cite{feys_toward_2025}.
The attack also leaves a cross-\gls{ap} fingerprint on the per-\gls{ap} $\beta_{l,k}$ that the \gls{gnn} can learn in the \gls{cf} scenario, unlike in a colocated one. Fig.~\ref{fig:gnn_arch} illustrates the proposed \gls{gnn} precoder $f_\theta$, which maps the spoofed estimate $\hat{\mat{H}}$ to the signal precoder $\mat{W}_{\mathrm{sig}}$ via a bipartite graph with $L$ \gls{ap} nodes, $K$ user nodes, and all $LK$ \gls{ap}-user pairs as edges; the \gls{gnn} output is later combined with the null-space \gls{an} block described below to form the \gls{gnn}-\gls{an} scheme evaluated in Section~\ref{sec:results}. Each \gls{ap}-user edge $(l,k)$ carries the raw feature $[\mathrm{Re}\{\hat h_{l,k}^{\mathrm{att}}\},\mathrm{Im}\{\hat h_{l,k}^{\mathrm{att}}\},\beta_{l,k}/\bar\beta_{\mathrm{b}}]$, where $\bar\beta_{\mathrm{b}}$ is the mean of $\{\beta_{l,k}\}$ across the batch and keeps the large-scale term at unit scale; a learnable map $\phi_{\mathrm{in}}$ projects and aggregates the edge features into initial node embeddings $\vect{a}_l^{(0)}$ and $\vect{u}_k^{(0)}$ of dimension $d$. Each of the $T$ layers then updates every node embedding by pooling messages from its neighbours,
\begin{equation}
    \vect{f}_i^{(t+1)} = \mathrm{UPDATE}\Bigl(\vect{f}_i^{(t)},\,\mathrm{AGG}\bigl(\{\vect{f}_j^{(t)}\}_{j\in\mathcal{N}(i)}\bigr)\Bigr)
    \label{eq:mp}
\end{equation}
where $\vect{f}_i^{(t)}\!\in\!\{\vect{a}_l^{(t)},\vect{u}_k^{(t)}\}$ is the layer-$t$ embedding of node $i$, and $\mathcal{N}(i)$ collects its neighbours in the bipartite graph, either the $K$ user nodes or the $L$ \gls{ap} nodes. Here AGG sums the neighbour messages, and UPDATE is a trainable network shared across nodes. Since every \gls{ap} node aggregates messages from all $K$ user nodes and every user node from all $L$ \gls{ap} nodes, information about the cross-\gls{ap} distortion covariance $\mat{C}_d$ propagates across cooperating \glspl{ap} through message passing, as shown in Fig.~\ref{fig:gnn_arch}. Each precoder entry $\tilde{w}_{l,k}$ is produced by the output layer $\phi_{\mathrm{out}}$ applied edge-wise to the pair $(\vect{a}_l^{(T)},\vect{u}_k^{(T)})$ of final \gls{ap} and user embeddings, so that permuting users permutes the columns of $\tilde{\mat{W}}=[\tilde w_{l,k}]$ identically and equivariance is preserved. The rows are then rescaled per \gls{ap} so that $\|\vect{w}_l^{\mathrm{sig}}\|^2=\rho\,P_l^{\max}$, where $\rho\!\in\!(0,1]$ is the signal power fraction. The remaining $(1-\rho)P_l^{\max}$ is allocated to a null-space \gls{an} block~\cite{goel_guaranteeing_2008,zhu_linear_2016}. Since one $\alpha_{\mathrm{an}}$ serves all \glspl{ap}, it is set to the smallest per-\gls{ap} value so no \gls{ap} exceeds its budget,
\begin{equation}
    \mat{W}_{\mathrm{an}} = \alpha_{\mathrm{an}}\,\mat{P}_{\mathrm{null}},
    \qquad
    \alpha_{\mathrm{an}} = \min_l\sqrt{\tfrac{(1-\rho)P_l^{\max}}{[\mat{P}_{\mathrm{null}}\mat{P}_{\mathrm{null}}^H]_{l,l}}}
    \label{eq:an}
\end{equation}
where $\mat{P}_{\mathrm{null}}\!\in\!\C^{L\times(L-K)}$ is an orthonormal basis of $\mathrm{null}(\hat{\mat H}^H)$ obtained via \gls{svd}. The precoder itself satisfies $\hat{\mat{H}}^H\mat{W}_{\mathrm{an}}=\mat{0}$; the residual leakage at legitimate users after the \glspl{pa} then arises from the two mechanisms discussed after~\eqref{eq:sindr_eve}. The final transmit vector is $\vect{x}=\mat{W}_{\mathrm{sig}}\vect{s}+\mat{W}_{\mathrm{an}}{\boldsymbol{\eta}}$, with ${\boldsymbol{\eta}}$ as defined in Section~\ref{sec:system_model}, independent of $\vect{s}$.
Following~\cite{zhu_linear_2016,atiya_secure_2024,zhang_secure_2020}, the closed-form baseline replaces the \gls{gnn} signal columns with the \gls{zf} pseudo-inverse scaled by a single common factor,
\begin{equation}
    \mat{W}_{\mathrm{sig}}^{\mathrm{ZF}}\!=\!\alpha_{\mathrm{zf}}\,\mat{Z},
    \ \
    \alpha_{\mathrm{zf}}\!=\!\min_l\sqrt{\tfrac{\rho P_l^{\max}}{[\mat{Z}\mat{Z}^H]_{l,l}}},
    \ \
    \mat{Z}\!=\!\hat{\mat H}(\hat{\mat H}^H\hat{\mat H})^{-1},
    \label{eq:zf}
\end{equation}
so that the per-\gls{ap} signal budget $\rho P_l^{\max}$ is respected at every \gls{ap}, and the common scaling preserves the zero-forcing property $\hat{\mat H}^H\mat{W}_{\mathrm{sig}}^{\mathrm{ZF}}=\alpha_{\mathrm{zf}}\mat{I}_K$ on the estimated channel. The identical null-space \gls{an} block of~\eqref{eq:an} is then appended. Both schemes therefore share the same power budget and the same \gls{an}. ZF-\gls{an} is the standard linear reference here~\cite{zhu_linear_2016,zhang_secure_2020,atiya_secure_2024} and sees the same \gls{pa} distortion without accounting for it and also a \gls{z3ro}-style precoder~\cite{rottenberg_z3ro_2023} is designed for the single-user case. The learned mapping $\mat{W}_{\mathrm{sig}}\!=\!f_\theta(\hat{\mat{H}})$, with trainable weights $\theta$, is trained by minimising the negative of the objective in~\eqref{eq:problem} directly on unlabelled channel samples,
\begin{equation}
    \theta^* = \arg\min_{\theta}\ -\E_{\vect{h}_e}\!\Bigl[\tfrac{1}{K}\!\textstyle\sum_{k=1}^{K}\!R_{\mathrm{sec},k}\bigl(f_\theta(\hat{\mat{H}})\bigr)\Bigr]
    \label{eq:loss}
\end{equation}
where $\E_{\vect{h}_e}[\cdot]$ is computed empirically by averaging over $N_{\mathrm{b}}$ mini-batch channel realisations from $N_{\mathrm{tr}}$ training samples, with $N_{\mathrm{val}}$ held out for validation, and $N_e$ eavesdropper picks per sample. The $N_e$ positions are uniform over the area, independent of the users. During training, the attacker's target index $p$ and pilot power $q_e$ are randomised per sample so the precoder sees varied attack conditions. The rates $R_{\mathrm{sec},k}=[\log_2(1+\gamma_k)-\log_2(1+\gamma_{e,k})]^+$ are computed from~\eqref{eq:sindr_leg}--\eqref{eq:sindr_eve}. The steps from the precoder $f_\theta(\hat{\mat{H}})$ through the polynomial \gls{pa}~\eqref{eq:pa_poly} to the \glspl{sindr}~\eqref{eq:sindr_leg}--\eqref{eq:sindr_eve} are differentiable in $\theta$, so backpropagation yields the gradient.

\section{Simulation Results}
\label{sec:results}
Table~\ref{tab:sim_params} lists the simulation parameters. The dataset is generated by placing user and eavesdropper positions over the deployment area once per experiment for a static deployment, computing large-scale gains from the ITU-R path-loss model normalised to per-user means $\bar\beta_k,\bar\beta_e$, and sampling independent small-scale coefficients per coherence block; the resulting large-scale fading of the legitimate user is known at the \glspl{ap} and is held fixed during training. The training uses a cosine learning-rate schedule from $10^{-3}$ to $10^{-4}$ and, per sample, randomises $q_e\!\sim\!\mathcal{U}[0.1,\,1.0]$ and, in the secrecy-rate loss, averages over $N_e$ Eve candidate locations selected per sample. The validation objective stops improving after the first few tens of epochs; later epochs change it by less than $0.1\%$. All outage rates reported below are computed on a held-out test set of $N_{\mathrm{te}}$ samples. The $\varepsilon$-outage secrecy rate $R^{(\varepsilon)}_{\mathrm{sec}}$ is the largest rate $R$ such that $\Pr\{R_{\mathrm{sec},k}\!<\!R\}\!\le\!\varepsilon$ across channel and attack realisations.

\begin{table}[t]
\centering
\caption{Simulation parameters.}
\label{tab:sim_params}
\renewcommand{\arraystretch}{1.0}
\setlength{\tabcolsep}{3pt}
\scriptsize
\begin{tabular}{@{}l l l@{}}
\toprule
\textbf{Symbol} & \textbf{Description} & \textbf{Value} \\
\midrule
$L,K$            & \glspl{ap}, users                              & $\{8,16,32\},\,3$ \\
Area, $f_c$      & Area, carrier freq.                            & $240^2\,\text{m}^2$, $3.5$ GHz \\
$\bar\beta_{k}$  & Mean legit.\ large-scale gain~\cite{noauthor_guidelines_nodate} & $0.7$ \\
$\bar\beta_{e}$  & Mean Eve large-scale gain~\cite{noauthor_guidelines_nodate}   & $0.5$ \\
$\sigma^2,\sigma_n^2$ & Uplink / downlink noise variance          & $0.1,\,0.1$ W\\
$P_t,\,P_l^{\max}$ & Total, per-\gls{ap} transmit power           & $\{8,16,32\}$ W, $P_t/L$ W \\
$\tau_p$         & Pilot length                                   & $5$ \\
$q_k,q_e$        & Legit., attacker pilot power (training)        & $1,\,\mathcal{U}[0.1,1]$ W \\
IBO, fit range   & \gls{ibo} range, poly.\ fit interval            & $[-15,0]$ dB, $[0,\,3\sqrt{P_l^{\max}}]$ \\
$\text{IBO}_{\mathrm{tr}}$ & Training back-off                    & $-1$ dB \\
$2N{+}1,\alpha_{\mathrm{ridge}}$ & Poly.\ degree, ridge          & $15,\,10^{-6}$ \\
$\rho$           & Signal power fraction                          & $0.8$ \\
$T,d$            & \gls{gnn} depth, width                        & $6,\,256$ \\
$N_e$            & Eve picks per sample                            & $25$ \\
$N_{\mathrm{tr}},N_{\mathrm{val}},N_{\mathrm{te}}$ & Training / validation / test set size & $350\text{k},\,50\text{k},\,16\text{k}$ \\
$N_{\mathrm{b}}$, lr & Batch size, lr schedule                    & $256$, $10^{-3}\!\rightarrow\!10^{-4}$ \\
$N_{\mathrm{ep}}$    & Training epochs                            & $145$ \\  
\bottomrule
\end{tabular}
\end{table}

\begin{figure}[t]
    \centering
    \includegraphics[width=0.7\linewidth]{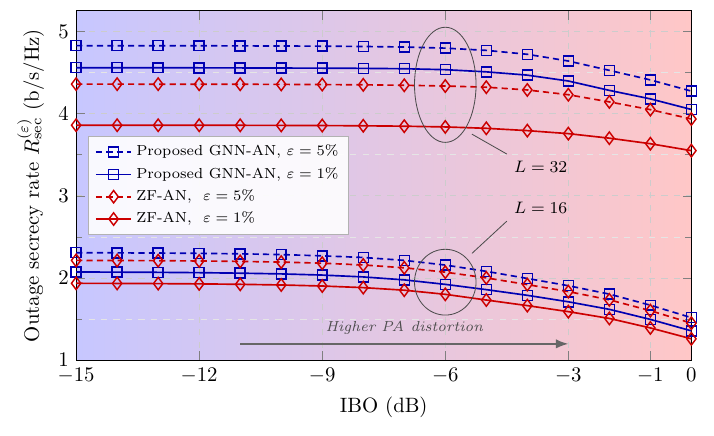}
    \caption{Outage secrecy rate versus IBO at $q_e/q_p=0.5$.}
    \label{fig:outage_rate_vs_ibo}
\end{figure}

\begin{figure}[t]
    \centering
    \includegraphics[width=0.7\linewidth]{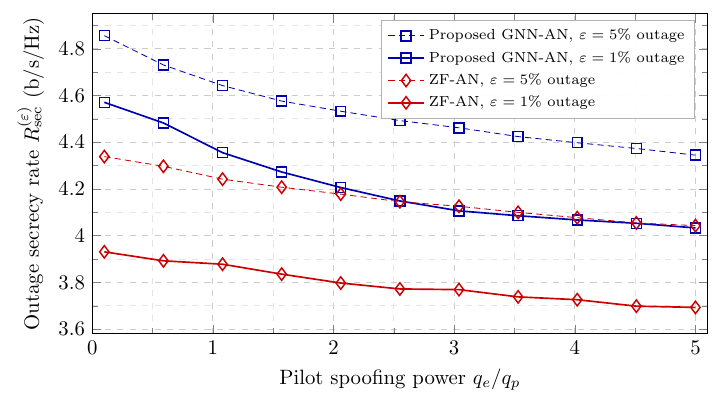}
    \caption{Outage secrecy rate versus $q_e/q_p$ at $\text{IBO}=-5$ dB.}
    \label{fig:outage_rate_vs_qe}
\end{figure}

\begin{figure}[t]
    \centering
    \includegraphics[width=0.7\linewidth]{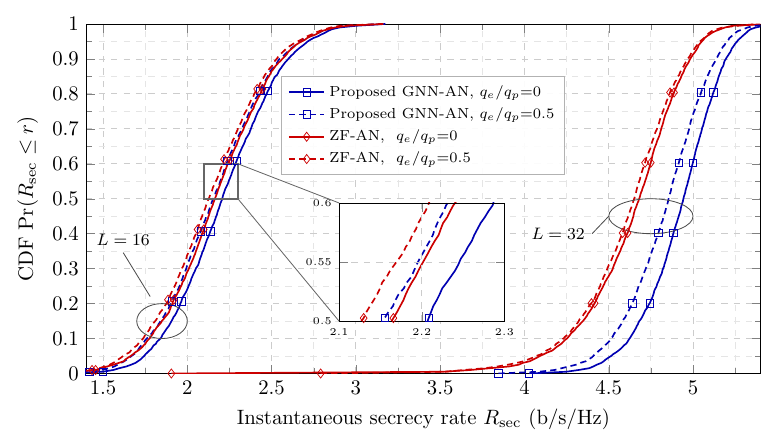}
    \caption{Empirical CDF of the secrecy rate at $\text{IBO}=-1$ dB.}
    \label{fig:cdf_secrecy}
\end{figure}

\begin{figure}[t]
    \centering
    \includegraphics[width=0.7\linewidth]{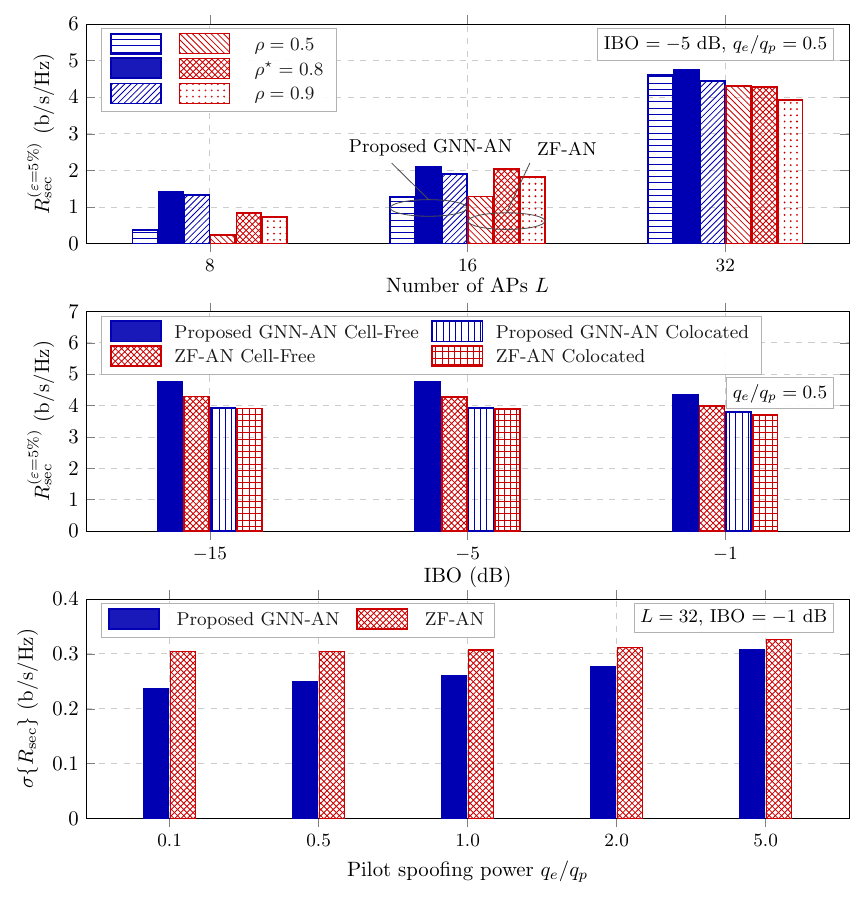}
    \caption{\gls{gnn}-\gls{an} vs \gls{zf}-\gls{an}. Top, $5\%$-outage secrecy rate vs $L$ and $\rho$. Middle, $L{=}32$ single-antenna \glspl{ap} for \gls{cf} vs a $32$-antenna \gls{bs} for colocated. Bottom, standard deviation of secrecy rate vs $q_e/q_p$.}
    \label{fig:outage_5pct_vs_rho_L}
\end{figure}

To assess secrecy performance from the linear to saturation regime of the \gls{pa}, Fig.~\ref{fig:outage_rate_vs_ibo} sweeps $\text{IBO}\!\in\![-15,0]$ dB at $q_e/q_p=0.5$. Both schemes lose rate as the \gls{pa} saturates, but \gls{gnn}-\gls{an} stays above ZF-\gls{an} throughout. At $L=32$ and $\text{IBO}=-1$ dB, \gls{gnn}-\gls{an} improves the $1\%$ and $5\%$ outage rates over ZF-\gls{an} by $15\%$ and $8.9\%$, and the gains stay above $14.1\%$ and $8.6\%$ across the full \gls{ibo} range. The gain therefore persists where the \gls{pa} is almost linear, and Fig.~\ref{fig:outage_5pct_vs_rho_L} shows it nearly vanishing in the colocated case, so it comes mostly from the per-\gls{ap} fingerprint.
Fig.~\ref{fig:outage_rate_vs_qe} fixes $\text{IBO}=-5$ dB and varies $q_e/q_p$ up to $5$, covering weak to strong pilot spoofing. Both schemes lose rate as $q_e/q_p$ grows because the bias in~\eqref{eq:mmse_att} scales with $q_e\beta_{l,e}$, but \gls{gnn}-\gls{an} stays above ZF-\gls{an} throughout, with gains of at least $8.9\%$ and $7.5\%$ at the $1\%$ and $5\%$ outage levels at $L=32$ across the full range, so one trained model generalises across attack strengths.
Fig.~\ref{fig:cdf_secrecy} shows the empirical \gls{cdf} of the instantaneous secrecy rate at $\text{IBO}=-1$ dB, $q_e/q_p=0.5$, and $\rho=0.8$, with and without attack. Both schemes shift left under attack; \gls{gnn}-\gls{an} stays to the right of ZF-\gls{an} in both cases, with the largest gap in the low-rate tail. The training loss randomises the attacker's target index $p$ per sample, so no user is consistently worse off.

\begin{figure}[t]
    \centering
    \includegraphics[width=0.7\linewidth]{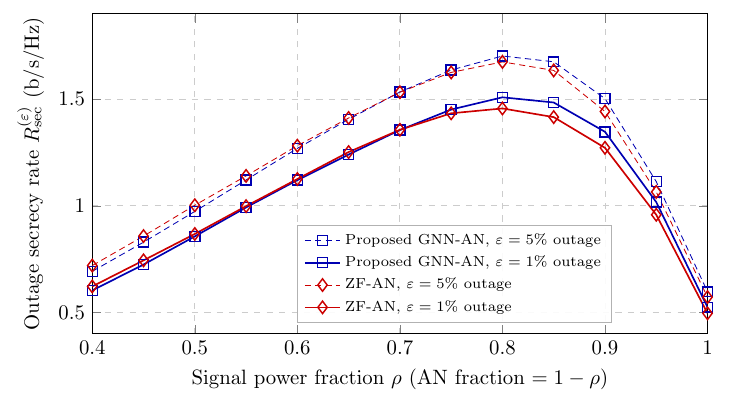}
    \caption{Outage secrecy rate versus signal power fraction.}
    \label{fig:outage_rate_vs_rho}
\end{figure}

To test robustness to the signal/\gls{an} power split, Fig.~\ref{fig:outage_rate_vs_rho} varies the signal power fraction $\rho$ from $0.4$ to $1.0$ under attack strengths of $q_e/q_p=0.5$ and $\text{IBO}=-1$ dB. Both schemes peak near the best signal fraction $\rho^\star\approx 0.8$, where user signal and \gls{an} power are balanced. At $L=16$, \gls{gnn}-\gls{an} improves the $1\%$ and $5\%$ outage secrecy rates over ZF-\gls{an} by $3.6\%$ and $1.7\%$ at $\rho=0.8$, growing to $5.9\%$ and $4.2\%$ at $\rho=0.9$ as the AN budget decreases.
Fig.~\ref{fig:outage_5pct_vs_rho_L} shows the \gls{gnn}-\gls{an} advantage over ZF-\gls{an} from three angles. The top subplot shows that the gap persists across $L\in\{8,16,32\}$. The middle subplot compares the \gls{cf} deployment with a colocated scenario with $L{=}32$ single-antenna \glspl{ap} for \gls{cf} versus one \gls{bs} of $32$ antennas, removing the per-\gls{ap} $\beta_{l,k}$ fingerprint that the \gls{gnn} exploits, so \gls{gnn}-\gls{an} reduces to \gls{zf}-\gls{an}. The bottom subplot shows that \gls{gnn}-\gls{an} also has a lower across-realisation standard deviation $\sigma\{R_{\mathrm{sec}}\}$, indicating a more stable secrecy rate.

\subsection{Complexity and Inference Time}
\label{subsec:complexity}
Table~\ref{tab:inference_time} reports the runtime per coherence block on an Intel Core Ultra~7~155H \gls{cpu} at a $256$-block batch. The \gls{gnn} forward pass costs $\mathcal{O}(T(LK+L+K)d^{2})$ \glspl{flop}, split into per-edge $\mathcal{O}(LKd^{2})$ projections and per-node $\mathcal{O}((L+K)d^{2})$ updates; the \gls{zf}-\gls{an} baseline is dominated by the null-space \gls{svd} at $\mathcal{O}(L^{3})$. Batched \gls{gnn} matrix multiplications are more hardware-efficient than the \gls{svd}, so the wall-clock ratio is smaller than the \gls{flop} ratio. \gls{gnn} training is offline and does not add to per-block cost.

\begin{table}[t]
  \centering
  \caption{Inference cost per coherence block at $L=32$.}
  \label{tab:inference_time}
  \footnotesize
  \setlength{\tabcolsep}{5pt}
  \begin{tabular}{@{}l c c@{}}
    \toprule
    Quantity & \gls{gnn}-\gls{an} & ZF-\gls{an} \\
    \midrule
    \glspl{flop}  & $\mathcal{O}(T(LK{+}L{+}K)d^{2})\!\sim\!5.2\!\times\!10^{7}$ & $\mathcal{O}(L^{3})\!\sim\!3.3\!\times\!10^{4}$ \\
    Throughput              & $1{,}500$ blocks/s & $5{,}700$ blocks/s \\
    Time        & $0.66$ ms & $0.18$ ms \\
    \bottomrule
  \end{tabular}
\end{table}
\section{Conclusion}
\label{sec:conclusion}
In a \gls{cf} network under active pilot spoofing, the nonlinear \gls{pa} distortion, usually treated separately by classical designs, is jointly handled by a single cooperative precoder that exploits both the distortion and the per-\gls{ap} fingerprint of the attack carried by the biased estimate. The gain over the ZF-\gls{an} baseline holds across the whole \gls{ibo} range, the whole attack-power range, and the whole signal-\gls{an} power split. Because the results assume a static deployment in which the trained model learns this specific geometry, future work will address mobile scenarios via geometry-conditioned inputs or per-deployment fine-tuning, attack detection and mitigation to neutralise spoofing before precoding, extension to colluding attackers and multi-antenna \glspl{ap}, a capacity limited fronthaul, and hardware validation.

\bibliographystyle{IEEEtran}
\bibliography{references/glb_references}

\end{document}